\documentclass[a4paper]{jpconf}
\usepackage{graphicx}
\begin{document}
\title{Time-scaled scenario of \\
low-energy heavy-ion collisions} 

\author{Yoritaka Iwata}

\address{Graduate School of Science, The University of Tokyo}
\ead{y.iwata@gsi.de}

\begin{abstract}
The underlying scenario of low-energy heavy-ion collisions is presented based on
time-dependent density-functional calculations.
A classification of several types of reaction dynamics is given with respect to their time-scales.

\end{abstract}

\section{Introduction}\label{sec1}
Heavy-ion reactions play an essential role in the evolution of our universe.
Indeed, our universe consists of nuclei, electrons and so on.
In particular most of the chemical elements (nuclei) are realised as final products of low-energy heavy-ion
reactions such as fusion reactions.
Therefore studies of the low-energy heavy-ion collision provide fundamental information about the chemical evolution of our universe.

In this paper, the reaction dynamics of heavy-ion reactions is discussed based on the time-dependent density-functional theory (TDDFT), with special attention to the time scales.
It was quite recently that we had a certain understanding of the underlying
scenario of low-energy heavy-ion reactions \cite{iwata-2010,iwata-2012}, even though it is rather limited to the early and the intermediate stages of heavy-ion reactions.
This issue is not only related to the answer to the fundamental question ``how do heavy-ion
reactions take place'', but also yields some constraints on both the
dynamics and the final products of heavy-ion reactions.

\section{Time scales} \label{sec2}
\subsection{Some energies decisive for heavy-ion reactions}

We consider collisions between two heavy ions (two nuclei),
\[ ^{A_1}Z_1 ~+~ ^{A_2}Z_2,  \]
where $Z_i$ and $A_i$ denote the proton and mass numbers of the two colliding nuclei, respectively.
Let us denote heavy-ion collisions with
an incident energy of a few MeV per nucleon as low-energy heavy-ion
collisions.
It corresponds to energies around or lower than the Fermi
energies of each colliding nucleus, so that low-energy heavy-ion reactions
mostly include fusion, fission, and fragmentation.

\begin{figure}[t]
\begin{center}
\includegraphics[width=12cm]{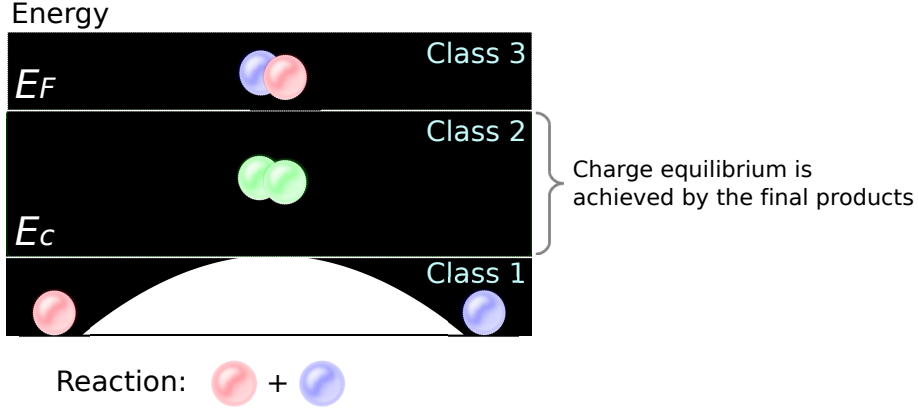} 
\caption{(Colour Online) Classification of heavy-ion collisions according to the incident energy.
There are three classes in low-energy heavy-ion collisions, where the colliding two heavy-ions are represented by red and blue spheres.
The colour of each sphere indicates the proton-to-neutron ratio.
Two nuclei with different proton-to-neutron ratios collide (in terms of the most probable time evolution) if the incident energy is higher than the Coulomb energy $E_C$.
For the energy region between $E_C$ and $E_F$, charge equilibrium is achieved by the final products (indicated by the green colour of the spheres), i.~e., an equilibrium with respect to proton-to-neutron ratio.
Another class is determined by the Fermi energy $E_F$.
For the definition of $E_C$ and $E_F$, see Eqs.~(\ref{eq1}) and (\ref{eq2}), respectively.}
\end{center}
\end{figure}

Let us begin with rough energy estimates.
The nucleus is a many-nucleon system in which there are two kinds of nucleons: protons and neutrons. 
Based on the mass of the nucleons, the typical energy scale for the nucleon is the order
of 1~GeV.
The nucleon is rather stable during the reaction process if the
incident energies are sufficiently smaller than 1~GeV per nucleon. 
It guarantees the stable existence of nucleons, as well as the validity of
treating low-energy heavy-ion collisions in the theoretical framework
based on the nucleonic degrees of freedom.
Meanwhile, there is another criterion energy; namely, the Fermi energy (denoted by $E_F$).
The Fermi energy is related to the stability and existence of the nucleus.
Considering the application to heavy-ion collisions, we derive
estimates for the Fermi energy in the centre-of-mass frame.
The kinetic energy of the projectile nucleus moving at the speed of
$x$~\% of the speed of light ($x < 50$) in the laboratory frame is
calculated to be
\begin{equation}
E_{\rm lab}/A_1=\frac{1}{2}m_N  \left(\frac{x}{100} c \right)^2  =\frac{1}{2}m_Nc^2 \left(\frac{x}{100} \right)^2=469 \left(\frac{x}{100} \right)^2  {\rm (MeV)}  
\end{equation} 
per nucleon, where $A_1$ denotes the mass of the projectile and the mass
of the nucleon was inserted as $m_N=938\,{\rm MeV}/c^2$.
Meanwhile, the centre-of-mass energy per nucleon is 
\[ E_x =  E_{\rm cm}/(A_1 + A_2) = (E_{\rm lab}/A_1) \frac{A_1 A_2}{(A_1+A_2)^2} = 469 \left(\frac{x}{100} \right)^2   \frac{A_1 A_2}{(A_1+A_2)^2}  ~{\rm (MeV)}.  \]
Consequently the Fermi energy per nucleon in the centre-of-mass frame is estimated as
\begin{equation} \label{eq1} E_{F} / (A_1 + A_2) = E_{x = 25} ~{\rm (MeV)},  \end{equation} 
where $x=25$~\% (of the speed of light) has been known as the typical value for the Fermi velocity of nuclei (for more accurate
estimates, see Eq.~(1) of \cite{iwata-2010}).
Here the Fermi energy per nucleon means the incident energy per nucleon which reaches the Fermi energy in collision situations; cf. the typical value of Fermi energy is equal to 40 MeV.
For example, in mass symmetric collision $A_1 = A_2$ with $E_{\rm lab}/A_1$ and $E_F/ (A_1 + A_2)$ are calculated to be 29.3~MeV and 7.3~MeV respectively.
In particular the latter value (7.3~MeV) is reasonable in comparison with the
results shown in Fig.~2 of \cite{iwata-2010}. 

In addition, the Coulomb energy ($E_C$) of the binary system, which is calculated by the
Coulomb energy between the two colliding nuclei, can be another criterion-energy in determining the
type of reaction dynamics.
Two nuclei cannot touch each
other if the total incident energies are less than the
Coulomb energy of the binary system, since the proton is a charged particle.
Using a simple approximation of the touching distance, the Coulomb energy per nucleon is estimated by
\begin{equation} \label{eq2} E_C/ (A_1 + A_2)  =  \frac {1.43987 ~ Z_1 ~ Z_2}{R~{\rm (fm)}}   \frac {1}{A_1 + A_2}    ~ \sim ~ \frac {1.43987 ~ Z_1 ~
  Z_2}{1.2~ (A_1^{1/3} + A_2^{1/3}) ~{\rm (fm)}}  \frac {1}{A_1 + A_2}   ~{\rm (MeV)}. \end{equation}
This energy is 1.7~MeV (per nucleon) for collisions between two identical $^{238}$U ($Z_1 = Z_2 = 92$
and $A_1 = A_2 = 238$), and 0.4~MeV (per nucleon) for collisions between two identical $^{4}$He ($Z_1 = Z_2 = 2$
and $A_1 = A_2 = 4$).
Therefore we see that the Coulomb energy is a few MeV per nucleon at most.

In low-energy heavy-ion reactions, there are two different criterion-energies.
The first one is the Fermi energy ($E_F$) determining whether the fast charge equilibration appears (lower) or not (higher)
(for the definition of the fast charge equilibration, see \cite{iwata-2010}).
The second one is the Coulomb energy ($E_C$) determining whether fusion appears (higher) or not (lower).
Depending on the energy, the reaction dynamics is classified into three types (Fig.~1).
For energy less than the Coulomb energy, two ions cannot have a contact.
For energy larger than the Coulomb energy and less than the Fermi energy, this energy region is characterised by the fast charge equilibration
phenomena, in which the proton-to-neutron ratio of each final product is
almost similar to that of the total system.
For higher energies, the fast charge equilibration does not appear, and the
proton-to-neutron ratio of each final product is not similar to that of the total system.
In these transitions of reaction types, the fusion is the concept completely included in the charge
equilibration. 
Note that, although it has been believed that the Fermi energy is decisive to whether multi-fragmentation
appears or not, multi-fragmentation
cannot generally take place at energies 5 to 10~MeV (per nucleon) higher
than the Fermi energy.  
On the other hand, fragmentation into a few pieces can appear even at
energies sufficiently less than the Fermi energy.

\begin{figure}[t]
\begin{center}
\includegraphics[width=14cm]{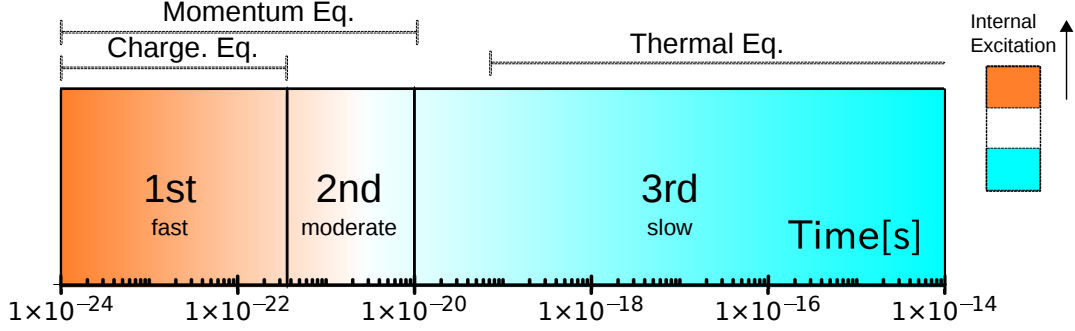}
\caption{(Colour Online) Time evolution of low-energy heavy-ion collisions.
Contact between the two ions takes place at 0~s.
Three typical time scales of low-energy heavy-ion collisions are illustrated
in which the first, second, and third period correspond to fast
charge, momentum and thermal equilibration, respectively.}
\end{center}
\end{figure}

\subsection{Time-scales in heavy-ion reactions}
The criterion-energies lead to the concept of time-scales of heavy-ion reactions.
\[ \hbar \omega = 0.65820 \times 10^{-21}~ \omega ~{\rm (MeV)}, \] 
where $\omega$~[s$^{-1}$] denotes the angular frequency, and the typical
duration time is estimated (as the time period) using $\omega= 2 \pi /T$.
Accordingly the typical time interval arising from the Fermi and the Coulomb energies
are $T_F =$1.0$\times$10$^{-22}$~s (the representative energy is 40~MeV) and
$T_C =$4.1$\times$10$^{-21}$~s (the representative energy is 1~MeV), respectively.  
Note that the typical duration time of low-energy heavy-ion collisions is
1000~fm/c ( 3.3$\times$10$^{-21}$~s).
In particular the Fermi energy is related to the motion of nucleons inside
nuclei, and more precisely to the fastest motion inside nuclei.
Such a motion is also related to the collective aspect of many nucleon
systems (cf.\ zero sound propagation \cite{iwata-2012}).
Meanwhile, the Coulomb energy is related to the relative motion of the two colliding nuclei.
For reference, the energy 1~GeV, which has been referred to as the
typical energy scale for the nucleon, corresponds to the time interval 4.1$\times$10$^{-24}$~s.  

From a different point of view there are three equilibria in classifying
heavy-ion collisions: charge, momentum and thermal equilibria.
The typical equilibration times of those three equilibria are
different from each other, and three different time-scales follow.
Let the three equilibration processes in low-energy heavy-ion
reactions be denoted by the fast, moderate, and slow processes (Fig.~2).
Typically the first period is associated with the fast charge
equilibration, which is the process at a time scale  of 10$^{-22}$~s.
The fast processes arise from the motion of nucleons inside a composite nucleus at the Fermi velocity.
The second period is associated with the momentum equilibration, which is the process at
a time scale of 10$^{-20}$~s. 
This time interval corresponds to the typical duration time of heavy-ion reactions.
The moderate processes arise from the motion of nucleons inside a composite
nucleus, where the composite nuclei formed in the first period do not correspond to stationary
states, therefore the nuclear and the Coulomb forces play a role in reducing
the corresponding imbalance.
The third period is associated with the thermal equilibration, which is the process taking
more than 10$^{-20}$~s.
The slow processes arise from the diffusion of nucleons.
This kind of motion gives rise to non-stationary time evolution even under charge and momentum equilibrium situations.

\subsection{Time scales of TDDFT calculations} 
TDDFT calculations are usually utilised to simulate events whose time scale is from 10$^{-23}$~s to 10$^{-20}$~s.
This utility of TDDFT calculations originally comes from the unit time step of calculations.
Indeed, by taking into account the accumulation of errors, we usually
have 10000 steps of calculations with a unit step of the order of 10$^{-24}$~s.
Note that the accumulation of errors should be more carefully treated for time-dependent calculations compared to the calculations for stationary states.
Consequently the present TDDFT calculations can only be applied to clarify the reaction processes whose time
scale is from 10$^{-23}$~s to 10$^{-20}$~s.
In this sense, based on the TDDFT calculations, it is difficult to
investigate reaction processes located on the third period in Fig.~2
without any special treatments.
In addition, the direct binary nucleon-nucleon collisions will play a
role in the long-term evolution, while they are mostly Pauli-blocked
in the initial phase.

The thermal effect cannot be fully treated in the TDDFT calculations, where the
total energy of the system is conserved during the calculation.
The iso-energetic time evolution is ultimately reduced to the unitary time
evolution in quantum mechanics, and the thermal effect treated in the TDDFT is only the energy
transfer between kinetic and potential energies, where the energy deposit in the potential energy is often called the internal excitation energy in the context of the TDDFT calculations.
Such energy transfers are only a part of possible energy transfers; indeed, even gamma-ray emission is not included.
It is worth mentioning here that the energy density functionals used in TDDFT are
described only by a small number of densities (for example, see \cite{iwata-2012b}), so that special time evolution such as heat flow, emissions and so on due to the localised distribution of internal excitation is highly suppressed in the TDDFT calculations.
Note that efficient cooling of composite nuclei is realized by particle emission. 

\section{Events in low-energy heavy-ion reactions} \label{sec3}
In the first period in Fig.~2, fast processes appear as consequences
of the fast charge equilibration.
It provides a strict limitation for the synthesis of exotic nuclei. 
Indeed, in case of fast charge equilibration, the proton-to-neutron ratios of the
final products are not far from the proton-to-neutron ratios of the the total system: $(Z_1 + Z_2)/(N_1 + N_2)$.
To avoid the fast charge equilibration, it is necessary to make the total reaction
time be shorter than 10$^{-22}$~s at the least, and the incident energy is required to be higher than $E_F$.
If the incident energy is higher than $E_F$, the reaction is already completed at the time 10$^{-22}$~s, and several fragments with different proton-to-neutron ratios appear (absence of fast charge equilibration). 

In the second period in Fig.~2, moderate processes appear as consequences of momentum equilibration.
For moderate processes, fusion, deep-inelastic reactions, or fragmentation into a few pieces takes place depending on the incident energy and the impact parameter.
All the moderate processes are affected by the fast process, so that most of the
fragments in low-energy heavy-ion collisions (whose incident energy is equal to a few MeV per nucleon) are in charge equilibrium.

In the third period in Fig.~2, slow processes appear as consequences of thermal equilibration.
For slow processes, fission, decay, and emission of particles are known.
Note that thermal transfer is related to fission and decay, while not essentially to
fragmentation into a few pieces.
All the slow processes follow after the fast and moderate processes.

\section{Conclusion}\label{sec3}
TDDFT can reveal the scenario of low-energy heavy-ion reactions, where TDDFT simulates only the most probable time evolutions.
Low energy heavy-ion collisions proceed step by step;
\begin{itemize}
\item in the early stage, fast charge equilibration inevitably appears mostly independent of the incident energy and the impact parameter; 
\item in the intermediate stage, fusion or fragmentation appears depending on the incident energy and the impact parameter;
\item in the late stage, fission, decay, or particle-emission appears.
\end{itemize}
Each time period corresponds to its specific equilibration dynamics: the fast charge, momentum and thermal equilibrations respectively.
In particular, charge and momentum equilibria are already achieved in the first and second periods, respectively.

\ack{This work was partially supported by the Helmholtz alliance HA216/EMMI.
Valuable comments from Prof. J. A. Maruhn are appreciated.
The author would like to express his sincere gratitude to Prof. T. Otsuka for perpetual encouragement.}

\end{document}